\documentclass[twocolumn,10pt]{asme2ej}

\usepackage{epsfig} 
\usepackage{graphicx}
\usepackage{float}
\usepackage{amsmath}
\usepackage{nccmath}
\usepackage{caption}
\usepackage{subcaption}

\title{Wing Expansion Menu - An approach for faster and more precise navigation with cascading pull-down menus}

\author{Manuel Zierl
    \affiliation{
    Email: manuel.zierl@web.de
    }	
}

\begin{document}

\maketitle    

%%%%%%%%%%%%%%%%%%%%%%%%%%%%%%%%%%%%%%%%%%%%%%%%%%%%%%%%%%%%%%%%%%%%%%
\begin{abstract}
This paper presents a new design suggestion for cascading pull-down menus to make user interaction with it faster and therefore easier: The Wing Expansion Menu (WEM). The proposal is based on the Steering Law \cite{accot1997beyond}, which implies a wider steering path for menu items. 
Our Approach combines this enlargement with a heuristic function that provides a probability with which the user will select an menu item. The menu can also be adapted to a wide variety of situations using certain variables. A user study of a WEM against a standard pull-down menu showed an average \textbf{improvement of 18.63\% }in user interaction speed. A second user study, which evaluated one of the significant innovations of the WEM compared to a similar approach, showed an average \textbf{improvement of 7.01\%} in user interaction speed.
\end{abstract}
\hspace*{3,6mm}\textit{\textbf{Keywords}: user menu navigation, steering law, cascading pull-down menus, wing expansion menu, WEM}

\section{Introduction}
\label{intro}
Pull-down menus have been established in UX design as one of the most commonly used methods for selection tasks. They use the beneficial fact that within a virtual system not all available functions must always be displayed. In addition, its tree-like structure helps to optimize/minimize the amount of choices according to Hick's Law \cite{hick1952rate} by dividing elements into hierarchically arranged submenus.

However, they have the disadvantage that they require careful cursor movement by the user. If the cursor leaves the desired path, another sub-menu opens up and a time delay arises in the interaction. In some cases the menu closes when the cursor leaves the menu interface. If this unintentionally happens, the entire steering task must be performed from the beginning. Other menus require an additional click outside the menu to close it. However, this is also not optimal as time is wasted, by this additional explicit step.

This paper discusses how cascading pull-down menus can be optimized in their form to make navigation easier for the user and therefore faster. First, we will briefly discuss related work dealing with similar topics.  We will then focus on steering law \cite{accot1997beyond}, which is a very good model for navigation tasks. Based on these concepts and two axiomatic assumptions we will make, we will introduce a new design suggestion for a pull-down menu: The Wing Expansion Menu (WEM).

Finally, we will evaluate our approach against a standard pull-down menu and show in an additional study that the new innovations of our approach were useful.

\section{Related Work}
\label{sec:Related Work}
The idea of modifying pull-down menus to ensure better performance for the user has already been discussed in other papers. For example, it has been shown that an additional artificial movement of the mouse cursor towards the submenu results in a significant increase in the speed with which elements were selected \cite{ahlstrom2005modeling}.

The idea of enlarging active areas of the menu, has also already been researched and approved because it is an obvious consequence of the Steering Law\cite{accot1997beyond}. Examples for these type of menus are the enlarged activation-area menus (EMUs)\cite{cockburn2006faster} or the adaptive activation area menus (AAMUs)\cite{tanvir2008aamu}. 

Steering law is a model designed to predict the time that a human needs to perform a two-dimensional steering task. The law can therefore be applied to navigation in a pull-down menu. Steering law basically predicts a shorter average time for a larger steering path.

The WEM is based on this idea and combines this enlargement with a heuristic function that provides a probability with which the user will select an menu item.

\section{Wing Expansion Menu}
\label{sec:Wing Steering Menu}
The approach of the WEM is based on two axiomatic assumptions we made about the human navigation in cascading menus that are intuitively true. First, the longer the user stays with the cursor on a menu item, the more likely it is that he wants to select this element. Secondly, the further to the right the user's cursor is on a menu item, the more likely it is that the user wants to select it. The second is only correct if we assume a cascading menu on the right side. For a menu that cascades to the left, the horizontal direction of this axiom must be inverted. However, since cascading menus on the right side occur more frequently, we will consider only those in this paper for simplicity. 

If we accept these two axioms as true, a meaningful menu should facilitate the selection of menu items which are more likely due to these axioms. The WEM builds on these two axioms, by making use of findings by the Steering Law that a wider steering path leads to a faster and therefore more precise selection.

\begin{figure}
    \centering
    \begin{subfigure}{0.18\textwidth}
        \includegraphics[width=0.8\linewidth]{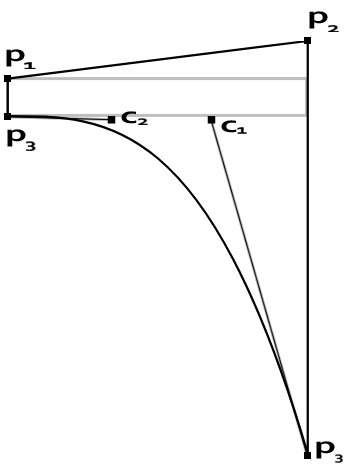}
        \caption{$\epsilon = 0$}
        \label{epsilon_0}
    \end{subfigure}%
    \begin{subfigure}{0.18\textwidth}
        \includegraphics[width=0.8\linewidth]{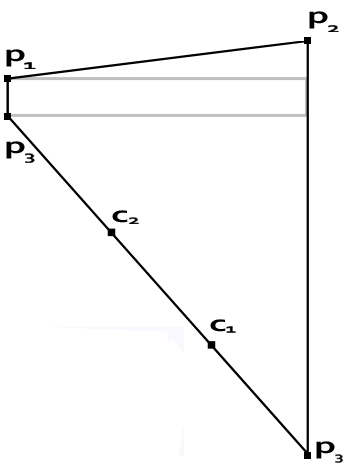}
        \caption{$\epsilon = 1$}
        \label{epsilon_1}
    \end{subfigure}
    \caption{The figures show a single menu item of the WEM with the variables: \\
    $ \eta = 1 \quad \quad   \alpha = 1 \quad \quad   \gamma = 10$.\\ 
    This means that the cursor of the user is in the maximum possible right-sided position and the element has 11 sub-items. The only difference between the two figures is the variable $\epsilon$.}
    
\end{figure}

Each menu item is defined in each state by a path between the four points in the given order:
\begin{ceqn}
\begin{align*}
p_1 \rightarrow  p_2 \rightarrow p_3 \rightarrow p_4 \rightarrow  p_1 
\end{align*}
\end{ceqn}

where $ p_1 \rightarrow  p_2 \rightarrow p_3 $ and $p_4 \rightarrow  p_1$ are connected by a straight line. And $p_3 \rightarrow p_4$ are connected by a bezier curve \cite{alt1997piecewise} with the two bezier curve handles $c_1$ and $c_2$. All points are defined by the relative (x,y)-coordinates as follows\footnote{We assume, as with digital systems usually, a coordinate system that has its origin in the upper left corner}:
\begin{itemize}
\item[-] $p_1 = (0,0)$
\item[-] $p_2 = (width, - \alpha *(heigth * \alpha * \eta )$
\item[-] $p_3 = (width, heigth + \alpha*(\gamma*heigth*\eta)-(heigth*\alpha*\eta)$
\item[-] $p_4 = (0, height)$
\item[-] $c_1 = (\frac{2}{3}, heigth + \alpha*((p_3.y-heigth)*\frac{2}{3})*\epsilon)$
\item[-] $c_2 = (\frac{1}{3}, heigth + \alpha*((p_3.y-heigth)*\frac{1}{3})*\epsilon)$
\end{itemize}
where the variables have the following meaning:
\begin{itemize}
\item[] \textbf{\textit{heigth}}: The height of a menu item in pixels.
\item[] \textbf{\textit{width}}: The width of a menu item in pixels.
\item[] $\boldsymbol{\eta}$: The horizontal position of the cursor in percent above a menu item, where the far left corresponds to 0 and the far right to 1. If $\eta$ is 0, so if the cursor is not over the element, the menu item simply has the shape of a rectangle with the dimensions $width$ x $height$. The further $\eta$ approaches the 1, the larger the area of the menu item becomes.
\item[] $\boldsymbol{\alpha}$: Value between 0 and 1, which controls the maximum size of a menu item. This is a static variable defined by the developer.
\item[] $\boldsymbol{\epsilon}$: Value between 0 and 1, which indicates the curvature of a menu item, where 1 means no curvature (s. Fig.  \ref{epsilon_1})and 0 means the greatest possible curvature (s. Fig.  \ref{epsilon_0}). This is a static variable defined by the developer.
\item[] $\boldsymbol{\gamma}$: number of child elements of the menu item -1.
\end{itemize}

Up to this point, we consider only the second axiom, which finds application through the variable $\eta$. However, it is also very important to include the first axiom. Otherwise, any menu item that the user only briefly touches at a right-sided position would lead to an immediate enlargement of this menu item. For this reason, WEM implements a time delay as a lower bound for opening a menu item. If the user is only briefly on a menu item, it is not opened.

\begin{figure}
\includegraphics[ width=0.5\textwidth]{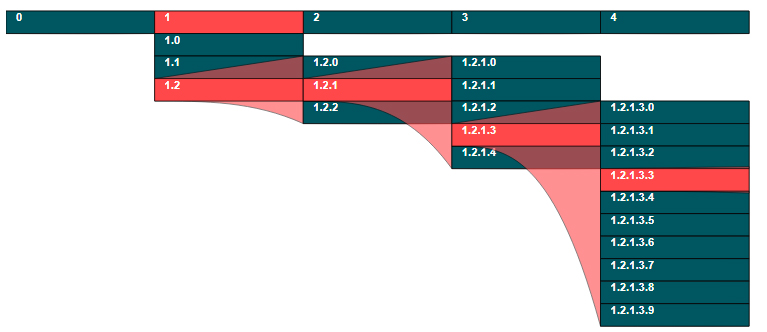}
\caption{The graphic shows a WEM ($\epsilon = 0$) that has been opened up to a depth of four submenus.}
\label{demo}
\end{figure}

\begin{figure}

    \centering
	\includegraphics[ width=0.3\textwidth]{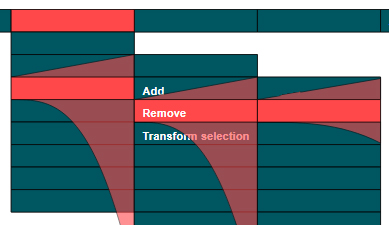}
	\caption{In this example, the text in one sub-menu is so large that it would be covered by the other opened menu-item. However, since the overlapping element is slightly transparent, it is still possible to read the text completely and also recognize the steering path.}
	\label{demo_opacity}
\end{figure}

Because opened menu items overlap other items on the same hierarchical level, it is possible that text is covered. This problem can be easily solved by displaying the overlapping area with a certain transparency (s. Figure \ref{demo_opacity}). Thus the user sees the steering path but can still read the text of other subitems.

A reduction of the variable $\epsilon$ results in a curvature and thus a reduction of the activation area. At first glance this would conflict with the Steering Law, but we must remember that the current menu item is only selected with heuristic probability. If the curvature is omitted, this makes it difficult to return to another item in the upper menu. Another reason why this curvature can be useful is because it better shows the user the relationship between his horizontal mouse position and the vertical height of the activation area. Because the relationship between these two variables is in any case quadratic independent of the variable $\epsilon$. For special cases, however, setting the variable $\epsilon = 1$ also allows the possibility for no curvature, which makes the WEM applicable for a wider range of use cases.

\section{Evaluation}
\subsection{Comparison with standard menu}
\label{Comparison with standard menu}
To find out how far using the WEM will affect the average time a user needs to find a predefined selection in a cascading pull-down menu we conducted a user study with 12 participants. A menu similar to that from Fig. \ref{demo} was presented to the participants. The labels consist of hierarchically sorted combinations of numbers. The user was then presented a certain combination of numbers which he had to find and select (click). The participants were divided into groups A and B. Each participant had to solve 16 of the described tasks. There were 6 tasks with the WEM and 6 without it\footnote{In reality, both were actually WEMs, but in the second case $\alpha$ was set to $0$, which implies that in no case a menu item enlarges}, which is the only change of the menu during the tasks. Group A started the first 6 tasks with the WEM and got then 6 tasks without it. For Group B it was the other way around. This methode was applied to minimize changes in the task duration caused by a possible learning effect. 

\begin{figure}
    \centering
	\includegraphics[ width=0.5\textwidth]{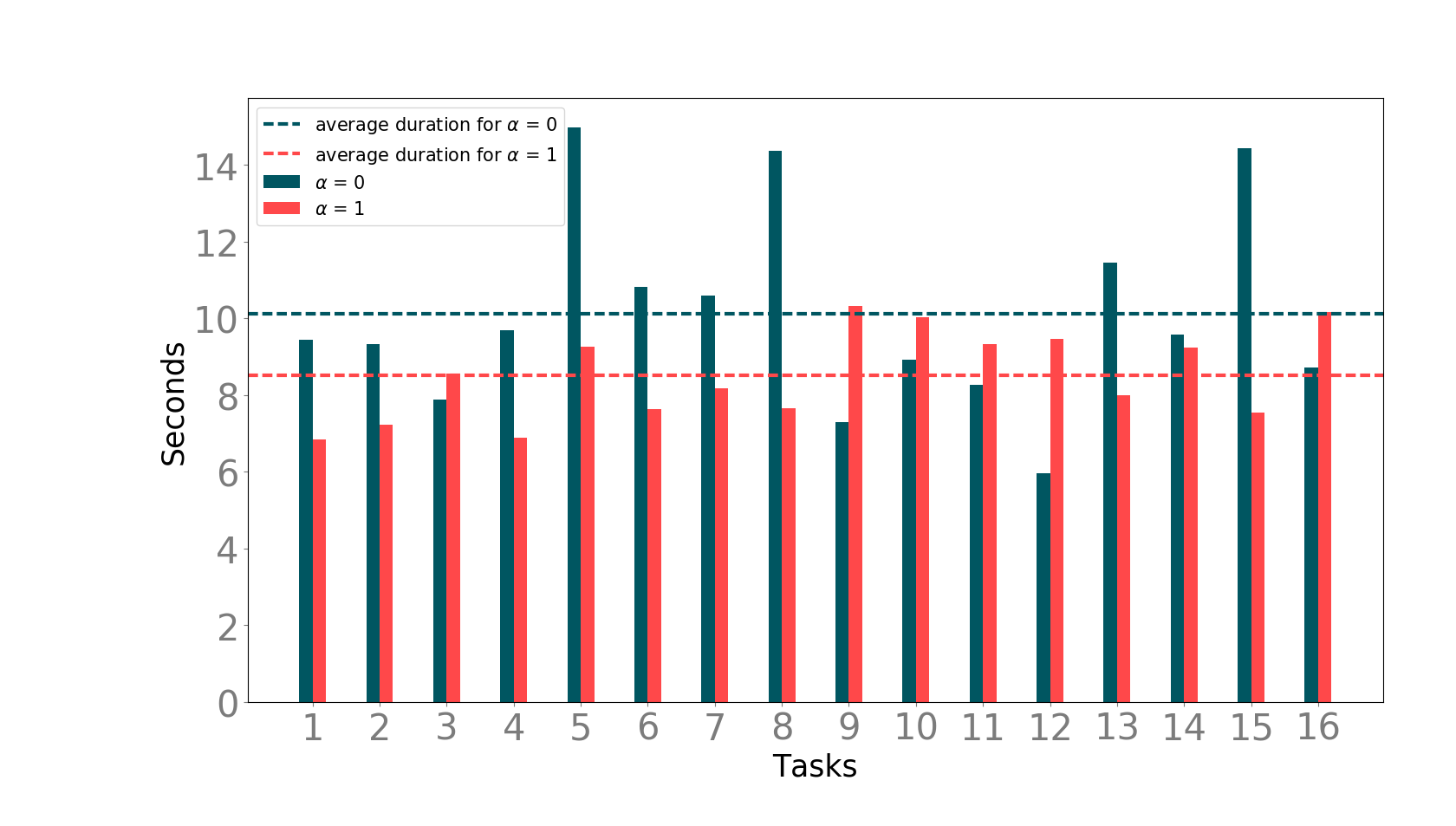}
	\caption{On the y-axis we see all 16 tasks on the x-axis is the average time in seconds needed to complete this selection task. Depending on the group, the tasks were to be solved for all participants in half with $\alpha = 0$ and in the other half with $\alpha = 1$.}
	\label{eval_01}
\end{figure}

This experiment showed that tasks could be executed faster by an average of \textbf{18.63\%} when using the WEM. Whereby one task without WEM took on average \textbf{10.10 seconds} and one with WEM \textbf{8.51 seconds}. Fig. \ref{eval_01} shows the average duration for all 16 tasks.

\subsection{Curvature analysis}
Since the WEM is based on the AMMU approach, we compared our approach in another user study with a corresponding menu. In order to change as few other variables within the study between the two menus, it is very useful that the WEM can be equated with such a menu by setting the variable $\epsilon = 1$ (that means no curvature), as this results in the triangular form demanded by the AMMU.

To do this, we have prepared a second study, which was carried out in almost the same way as in section \ref{Comparison with standard menu}. The only difference was that instead of changing the variables $\alpha$ in the two partial tasks from 1 to 0, the variable $\epsilon$ was changed\footnote{It should be noted here that with $\epsilon$ it behaves exactly the other way around as with $\alpha$. In this case $\epsilon = 0$ means our approach and $\epsilon = 1$ means the other approach}. In the study also 12 participants attended.

\begin{figure}
    \centering
	\includegraphics[ width=0.5\textwidth]{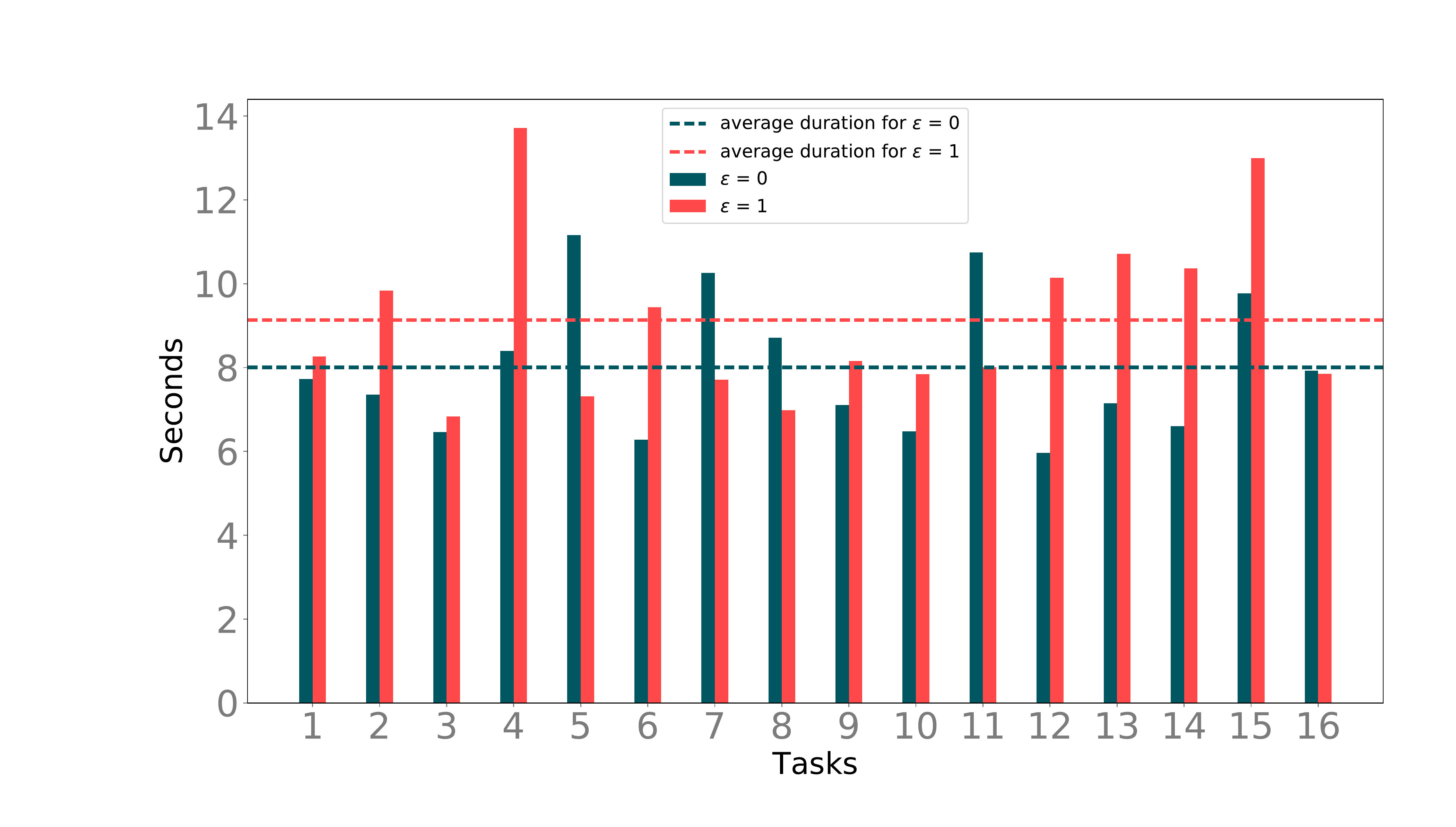}
	\caption{On the y-axis we see all 16 tasks on the x-axis is the average time in seconds needed to complete this selection task. Depending on the group, the tasks were to be solved for all participants in half with $\epsilon = 0$ and in the other half with $\epsilon = 1$.}
	\label{eval_02}
\end{figure}

The experiment showed that tasks could be executed faster by an average of \textbf{7.01\%} when using the curvature of the WEM. Whereby one task with triangular shape took on average \textbf{9.66} and one with curvature \textbf{8.98 seconds}. Fig. \ref{eval_02} shows the average duration for all 16 tasks.

\section{Conclusion}
In this paper the Wing Expansion Menu a new approach for a pull-down menu was presented, which should accelerate the navigation with a menu for the user and thus make interaction easier. The paper provided a precise formula for how this menu can be put together and discusses which possibilities of individual customization the menu provides through certain variables. Two user studies have shown that the menu has both an advantage over a standard menu ownership, and that it has advantages over the approaches from related work it is builds on.

\section{Future Work}
Although our evaluation has already shown that there is an improvement in the interaction with the WEM, it would be useful in the future to carry out a more detailed analysis and comparisons in different situations in order to obtain a more precise result on the exact effects.

Also in our paper a menu opening on the right side was assumed. It would be interesting to analyze a menu that opens on the left or alternately (depending on the space on the screen).

It might also be interesting to combine WEM with existing methods that accelerate interaction with pull-down menus (e.g. with the force-field approach \cite{ahlstrom2005modeling}).

\bibliographystyle{asmems4}
\bibliography{literature}  

\end{document}